\documentclass[review,authoryear,10pt]{elsarticle}

\usepackage{lineno,hyperref}
\modulolinenumbers[0]
\usepackage{fullpage}
\usepackage{amsmath}
\usepackage{amssymb}
\usepackage{longtable}
\usepackage{booktabs}
\usepackage{multirow}
\usepackage{diagbox}
\usepackage{soul}
\usepackage{amsmath}
\usepackage{enumitem}  
\usepackage[labelfont=bf]{caption}
\usepackage{placeins}
\usepackage{subcaption}
\usepackage{verbatim}
\usepackage[table,xcdraw]{xcolor}
\usepackage[colorinlistoftodos,prependcaption]{todonotes}
\usepackage{regexpatch}
\makeatletter
\xpatchcmd{\@todo}{\setkeys{todonotes}{#1}}{\setkeys{todonotes}{inline,#1}}{}{}
\makeatother
\journal{ArXiv}
\usepackage[utf8]{inputenc}

\usepackage{titlepic}
\usepackage{graphicx}

\begin{document}

\begin{frontmatter}

\title{\includegraphics[width=1.5cm]{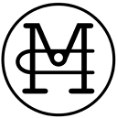} \\ MobilityCoins \\ A new currency for the multimodal urban transportation system}
%\tnotetext[mytitlenote]{Fully documented templates ae available in the elsarticle package on \href{http://www.ctan.org/tex-archive/macros/latex/contrib/elsarticle}{CTAN}.}

\author[mymainaddress]{Klaus Bogenberger\corref{mycorrespondingauthor}}
\cortext[mycorrespondingauthor]{Corresponding author \\ E-mail address: klaus.bogenberger@tum.de}
\author[mymainaddress]{Philipp Blum}
\author[mymainaddress]{Florian Dandl}
\author[mymainaddress]{Lisa-Sophie Hamm}
\author[mymainaddress]{Allister Loder}
\author[mymainaddress]{Patrick Malcolm}
\author[mymainaddress]{Martin Margreiter}
\author[mymainaddress]{Natalie Sautter}

\address[mymainaddress]{Chair of Traffic Engineering and Control, Department of Mobility Systems Engineering, School of Engineering and Design,Technical University of Munich, Germany}


\end{frontmatter}

%\linenumbers

\section{The problem}

Despite the continuous development in low-carbon technology and regulations of governments, transport greenhouse gas (GHG) emissions continue to increase globally which leads to enormous global challenges in regard to environment, society, and the city, and thus the mobility system. Transportation causes external costs, of which the climate change impacts are currently considered the most serious ones. \newline 
The Paris Agreement set the frame conditions for the mitigation of GHG emissions: reduction by around 80\% until 2050 compared with the 1990 level, which, in terms of continued normal development, raises to 95\%. But in fact, the road map until 2030 allows transportation to increase GHG emission by 8\% compared with 1990 and therefore doesn't send a clear signal towards industry and consumers to change neither technology nor behavior. This means, without serious mitigation policies being implemented, transport emissions could increase at a faster rate than emissions from other energy using sectors and reach around 12 Gt CO2eq/yr by 2050 \citep{IPPC}. In particular, cars with combustion engines are major polluters, accounting for 60.7\% of the total CO2 emissions from road transport in Europe \citep{EEA}. For this reason, incentivizing a shift to more environmentally friendly modes is a key step towards reaching the goals set out in the Paris Climate Agreement.

With ongoing urbanization and population growth, mobility systems face ever increasing challenges and requirements, especially given that urban space is limited, where the UN defined some of these challenges addressing sustainable development goals \citep{un2021}. In this context, the disproportionately large amount of space required by cars raises the question of the equitable and fair distribution of urban space today.
 Transport modes differ in the amount of emissions they produce and space they consume. However, in many countries, large portions of infrastructure costs are spread equally on all residents, regardless of their mobility behavior. This is leading to the problem of unequal internalization of negative externalities (i.e., a violation of the “polluter pays principle”). Additionally, revenues from fuel taxes are likely to decrease in the future due to the ongoing trend towards low- and zero-emission vehicles in the automotive sector, thus causing a revenue gap in public finances. The mobility system of the future must therefore consist of a fair and equitable infrastructure funding scheme. 
Historically, many traffic management approaches have failed due to lack of acceptance (e.g. a perceived “culture of prohibitions”) or a failure to provide successful pull-measures. Many existing traffic management instruments focus solely on one transport mode (e.g., parking pricing or public transport subsidies), or one problem (e.g. congestion pricing, speed limits). No existing measures seek to simultaneously influence both the supply and demand of multimodal transportation. Instead, small successes of such measures are all too often immediately undermined by induced traffic or large modal shifts to the improved system. Ambitious objectives regarding control of climate change require a broad set of instruments based on the same cost strategy and nonetheless challenging to generate sufficient incentives to change technology and behavior. \newline
Moreover, there is currently little opportunity for public participation in the decision-making process and design of mobility infrastructure. As of today, public participation in urban areas often lacks transparency, impact, and provides poor freedom of choice regarding the design and infrastructure of the mobility system.
To reach the goals of the Paris Climate Agreement, and address the aforementioned challenges, we propose a novel, innovative all-encompassing system to manage multimodal mobility in metropolitan areas:

\section{The idea}

\begin{figure}
    \centering
    \includegraphics[width=0.8\textwidth]{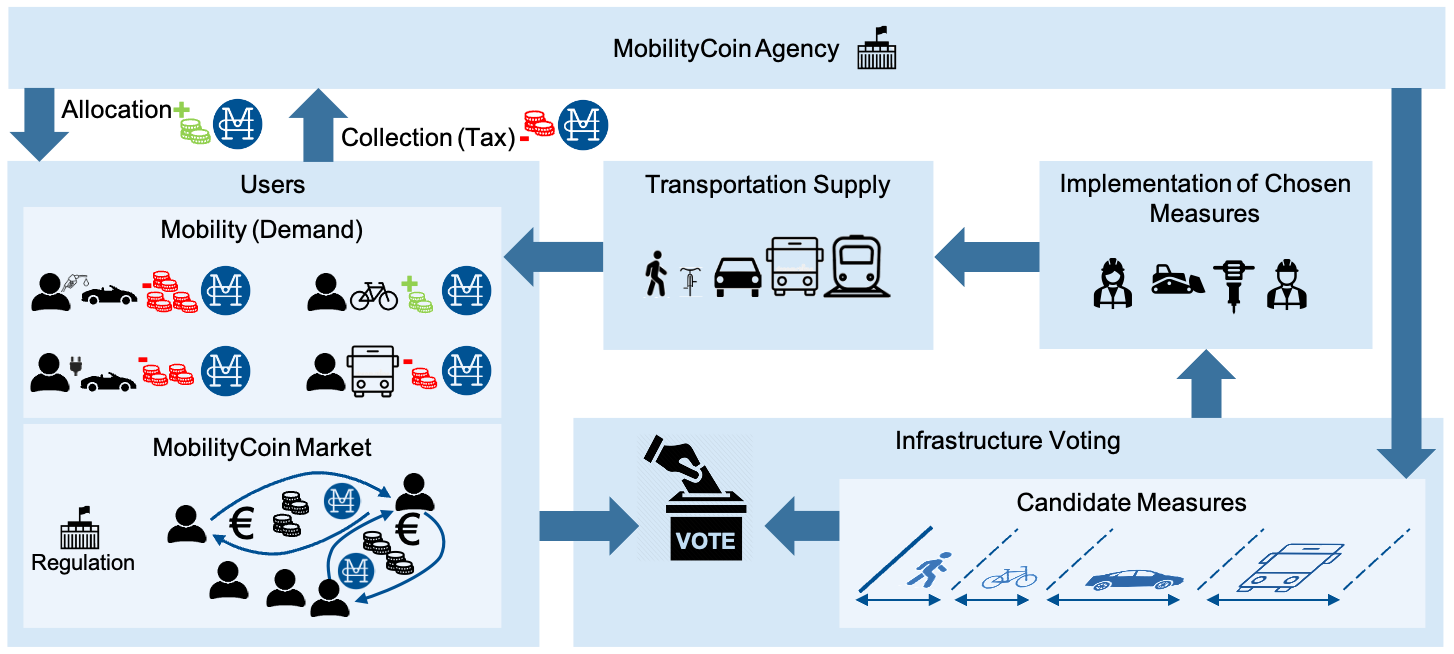}
    \caption{The MobilityCoin System}
    \label{fig:figure1}
\end{figure}

The MobilityCoin (MoC) is a new currency for paying for all your daily trips within a metropolitan area. At the beginning of each year, the MobilityCoin Agency allocates a specific number of coins to each person. The price of a trip depends on the mode, traffic state, occupancy, and trip length or duration. If you make a trip with a vehicle or public transport you pay a dynamic price with MobilityCoins. The use of environmentally friendly modes like biking or walking is incentivized by earning MobilityCoins. All users are allowed to buy and sell MobilityCoins in the MobilityCoin Market, which is regulated by the MobilityCoin Agency. It controls the market volume and price limits, it limits the number of coins a person can buy or sell and defines a transaction fee (see figure \ref{fig:figure1}). 

At the end of the year, remaining MobilityCoins can be used for infrastructure voting. Infrastructure voting empowers the public to re-allocate public space via infrastructure projects such like the construction of new dedicated bike lanes or improved public transport services. Each user’s voting power depends on their year-end remaining MobilityCoin balance, which incentivizes sustainable modes of transport.
The MobilityCoin system is an all-in-one approach, uniting different traffic management measures – it simultaneously optimizes supply and demand of the transportation system to accomplish environmental, social, and economic objectives, e.g., emission reduction, livability of urban space, equity, public participation, and infrastructure funding. The system influences infrastructure supply, mobility tool ownership, and mode choice. It is open to all types of technology, enables the fair financing of new infrastructure, and provides an adjustable, market-based and reliable pricing mechanism for mobility in limited urban space. The MobilityCoin system creates a transparent decision-making process and gradually re-allocates mobility space by enabling public participation. It also accounts for social acceptance and fairness by allocating a mobility budget to each person according to their mobility needs rather than their income or wealth.

\begin{figure}
    \centering
    \includegraphics[width=0.8\textwidth]{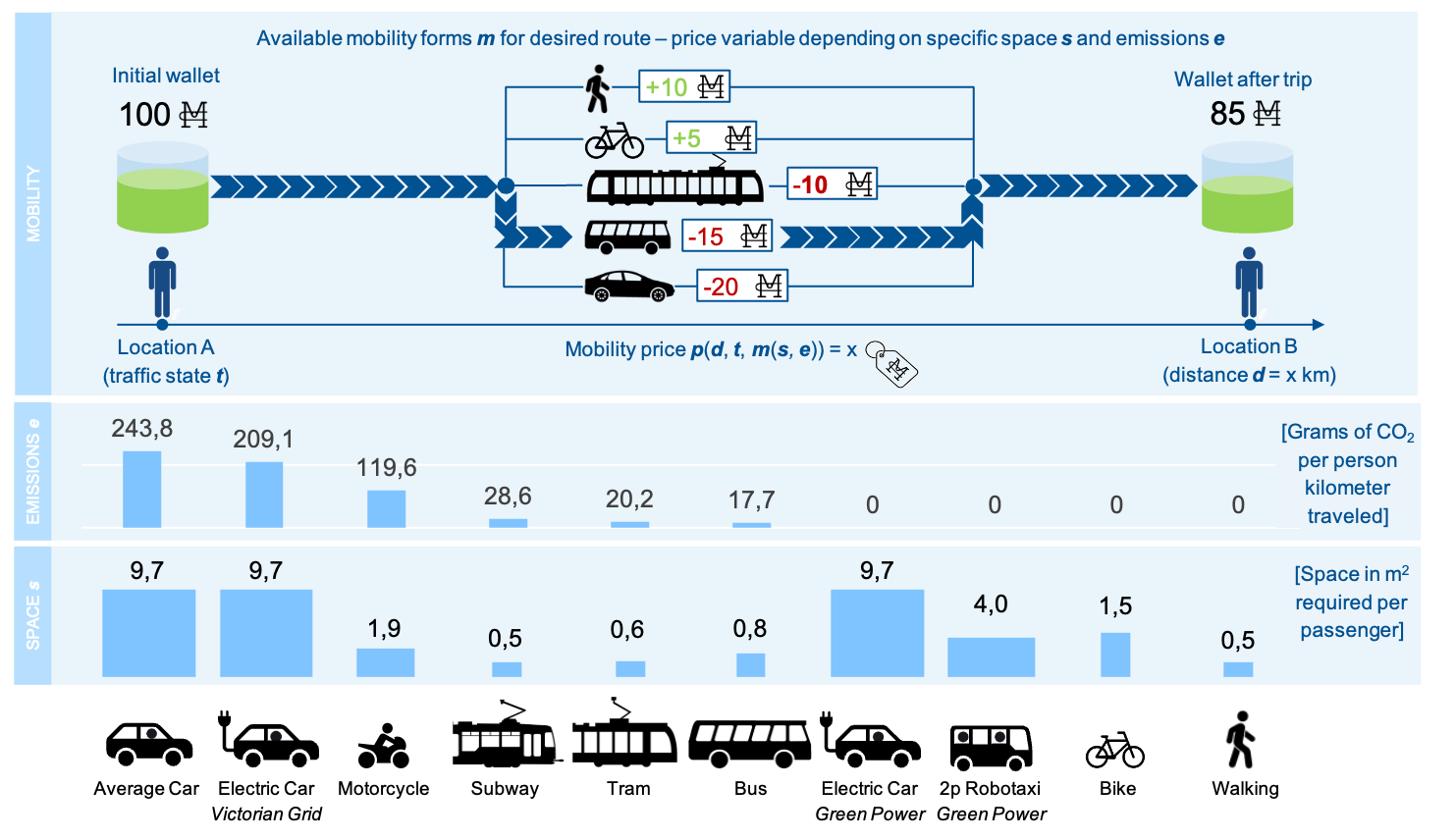}
    \caption{Mobility example, space consumption and emission production of different modes of transport}
    \label{fig:figure2}
\end{figure}

The main idea of the MobilityCoin System is that each person has an initial budget that they can use for their trips (see figure \ref{fig:figure2}). An individual’s allocated budget is based on factors such as mobility needs and the number and types of modes available in their area. Each mode has a dynamic price that depends on several factors, such as distance, emissions, and the current traffic state. The traveler makes a pre-trip decision based on this price. The change in wallet balance due to this price can range from highly negative, for example for single-vehicle rush-hour trips, to positive, for example for bike rides and walks. In other words, users can earn MobilityCoins by choosing environmentally friendly modes. In the example, the traveler has five options and ultimately chooses the bus trip that costs 15 MobilityCoins.
The core behavioral dimension of the proposed MobilityCoin system is that travelers respond to price signals following Marshall’s market equilibrium principle: demand for mobility decreases with higher MobilityCoins prices and vice versa (for incentives). This simple mechanism allows one to control and optimize the modal split and to incentivize sustainable transportation. 
%Reimbursement partially oder full
The MobilityCoin system is based on the polluter pays principle. The travelling person must pay the specific amount of MobilityCoins related to the externalities caused. The flow of MobilityCoins depends on the nature of the trip and the respective mode. From user's perspective, it can be positive for environmental friendly modes causing less externalities and negative for modes causing higher externalities (e.g. one passenger vehicle usage with comparably high specific GHG emissions and space consumption). On the earning side of the system there will be a threshold of daily earnable MobilityCoins implemented since the system design is meant to regulate traffic rather than generate the same. If a user has insufficient MobilityCoins for a ride, he is able to acquire new MoC's on the MobilityCoin Market for the current market price. \newline
Concerning the trip to work, different alternatives are available: (i) the employee works from home and gets an allowance per day from his employer for saving space, traffic capacity and as a result GHG emissions; (ii) the employee must pay MobilityCoins for the commute, however the employer reimburses the MobilityCoins as a sort of job ticket; (iii) the employee must pay for the trip to work as with a private trip without reimbursement; and (iv) the employee chooses an environmental friendly mode for getting to work, earns MobilityCoins while not getting charged by the MobilityCoin Agency. For case (i), (ii) and (iv) the employer must stay solvent for reimbursing the employees by buying additional MobilityCoins on the MobilityCoin market. Since business trips are demanded by work, they will be reimbursed mandatorily by the employer in analogy with case (ii) above. \newline

\begin{figure}[h]
    \centering
    \includegraphics[width=0.9\textwidth]{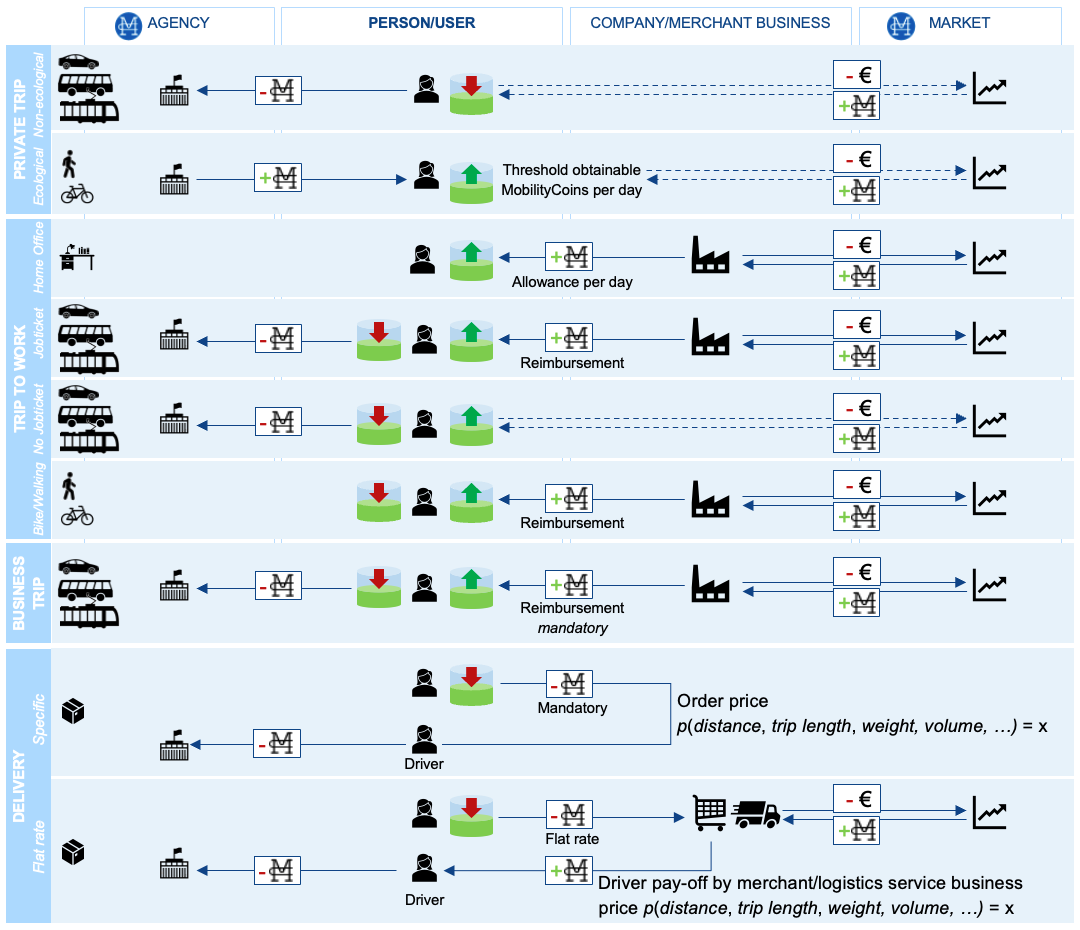}
    \caption{Flow of MobilityCoins for different scenarios}
    \label{fig:figure3}
\end{figure}

Deliveries and commercial transport in general generate a considerable amount of traffic in metropolitan areas \citep{EricHannon2016}. Within the MobilityCoin system, two conceivable approaches to cover deliveries are displayed in \ref{fig:figure3}: (i) the customer (polluter) pays a specific MobilityCoin delivery price as a function of variables affecting externalities (like distance, trip length, weight, volume, ...); or (ii) the customer pays a flat rate to the merchant business respectively logistics service provider that has to pay the specific MobilityCoin delivery price to the MobilityCoin Agency via the driver. In both cases the delivery price is sensitive to externalities caused and can be adjusted towards a mitigation of GHG emissions.    

As the system aims at meeting the overall environmental, social, and economic targets of the Paris Agreement, the MobilityCoin Agency can gradually increase or reduce the supply of MobilityCoins in order to reach said targets. For example, the total number of available MobilityCoins could be determined by the difference between the current system state (for example the current modal split) and the environmentally optimal modal split. This market intervention has an impact on the market outcome, eventually impacting the population’s mode choice. The MobilityCoin Agency regulates the market via the establishment of minimum and maximum prices, penalties, exchange fees, and other mechanisms.
All people with MobilityCoins remaining at the end of each year are eligible to vote on infrastructure measures, with their voting power being proportional to their MobilityCoin wallet balance. On the local level, such measures could be determined for example by crowd sourcing, in which voters can split their votes among multiple infrastructure measures according to the importance they assign to them. Infrastructure measures with the most votes would then be implemented. Another option is to vote on bundles of different infrastructure measures, so that splitting votes is not possible, but the voting process is simplified.
This public participation in the decision-making process creates an individual incentive to behave more sustainably, continuously develops infrastructure according to local will, and contributes to social acceptance of the mobility system. In contrast to infrastructure projects that are managed on the city- or regional level and which are decided upon by planners and policy-makers to ensure city-wide accessibility and connectivity, residents on the local neighborhood level are empowered to decide upon local, small-scale infrastructure adjustments. Measures can include the redistribution of space between modes, the development of new public transport lines, but also repurposing space as green space, restaurants, retail, or other uses. Operational changes could also be envisaged, for example higher service frequencies or capacities of public transport. However, these measures rely heavily on engineering and transport planning expertise, making the pre-selection of candidate measures in these cases crucial.
As described above, the MobilityCoin concept integrates several familiar policy and traffic management instruments, most of which can be classified primarily as economic instruments, however infrastructure voting can also be seen as an information-based instrument \citep{Axsen2020}. Given the high risk of completely revolutionizing regulation in the transport sector, and the high reward of reaching the goals of the Paris Climate Agreement with a sustainable and equitable instrument, a thorough investigation of details of the MobilityCoin concept is essential.

Basically, it can be possible to control traffic with a combination of pricing and rationing in order to improve the network state for every user due to the institution of a policy which recognizes people's personal needs by giving the chance to chose the form of the penalty they must pay for using a bottleneck \citep{Daganzo1995}. The idea of tradeable mobility permits \citep{Dogterom2017,Grant-Muller2014a} – i.e. a cap-and-trade system for mobility - traces back to Coase \citep{Coase2007}. Despite externalities, an efficient market outcome only results when few parties are involved, property rights are clearly defined, and little to no transaction costs exist, none of which is the case in mobility. The original idea of tradeable permits (TP) \citep{Dales1968} was transferred to the fields of transport \citep{Raux2005} and airport capacity allocation \citep{DeWit2008}. \cite{Verhoef1997} was then among the first to propose TPs for road traffic management. In theory, the mobility TPs proved successful in achieving a congestion reduction goal \citep{Yang2011}, and could also help to meet climate targets \citep{Musso2013}. As mobility TPs have an artificial market, their design may impact the final equilibrium \citep{Nie2012} and thus their ability to reach the target. Basic public support for a mobility TP scheme is found to be 25\%-30\% (similar to that for road pricing), with those opposing it generally not favoring any similar policy instrument \citep{Kockelman2005,Krabbenborg2021}.

\section{Outlook}

 Nevertheless, Krabbenbourg et al. (2021) notice that a mobility TP is yet far from applying to our current mobility system, while at the same time it is mentioned as a promising (theoretical) instrument \citep{Chen2021,Liu2020}. \newline
 In order to get closer to realization further research is required, especially when it comes to the infrastructure voting. Experiments and empirical studies can show how an involvement of users can influences social acceptance of such instruments.

\nocite{Dandl2021}
\nocite{Bracher2017}
%Auf nächste Seite
% References
%Referenzen Klaus 
    %Dandl regulating mobility on demand services IEEE
    %Bracher Modelling long term effects/pricing (2 paper)
    %

\bibliography{references}

\end{document}